\documentstyle[fleqn]{tp}

\input epsf

\def\bi{\begin{itemize}}
\def\ei{\end{itemize}}

\def\thedemobiblio#1{\smallskip\par
 \list{}{\labelwidth 0pt \leftmargin 1em \itemindent -1em \itemsep 1pt}
 \small \parindent 0pt
 \parskip 1.5pt plus .1pt\relax
 \def\newblock{\hskip .11em plus .33em minus .07em}
 \sloppy\clubpenalty4000\widowpenalty4000
 \sfcode`\.=1000\relax}

\input epsf
\let\sec=\section
\let\ssec=\subsection

\newcount\japif
\def\japeqn{\ifnum\japif=1
\begin{equation}\global\japif=0 \else
\end{equation}\global\japif=1\fi}

\def\japref{\item}
\def\gs{\mathrel{\lower0.6ex\hbox{$\buildrel {\textstyle >}
 \over {\scriptstyle \sim}$}}}
\def\ls{\mathrel{\lower0.6ex\hbox{$\buildrel {\textstyle <}
 \over {\scriptstyle \sim}$}}}
\newcount\equationo
\def\m@th{\mathsurround=0pt }
\def\eqalign#1{\null\,\vcenter{\openup1\jot \m@th
 \ialign{\strut\hfil$\displaystyle{##}$&$\displaystyle{{}##}$\hfil
 \crcr#1\crcr}}\,}

\def\kms{\;{\rm km\,s^{-1}}}

\def\hompc{\,h\,{\rm Mpc}^{-1}}
\def\mpcoh{\,h^{-1}\,{\rm Mpc}}

\def\ss{\scriptscriptstyle\rm}

\def\japfig#1#2#3#4#5#6{
\begin{figure*}
\centering\mbox{\epsfxsize=0.9\hsize\epsfbox[#1 #2 #3 #4]{#5}}
\caption[]{#6}
\end{figure*}
}

\def\apj{ApJ}
\def\apjs{ApJS}
\def\mn{MNRAS}

\begin{document}

\twocolumn[
\title{Models for large-scale structure}
\author{J.A. Peacock \\
{\it Institute for Astronomy,  University of Edinburgh} \\
{\it Royal Observatory, Edinburgh EH9 3HJ, UK}
}

\vspace*{16pt}   

ABSTRACT.\
This paper reviews selected aspects of the growth of
cosmological structure, covering the following general
areas: (1) expected characteristics of linear density
perturbations according to various candidate theories
for the origin of structure; (2) low-order theory for
statistical measures of fluctuations; (3) formation of
nonlinear structures and nonlinear evolution of the mass 
distribution; (4) the relation between the density
field and the galaxy distribution; (5) constraints
on cosmological models from galaxy clustering and
its evolution.
\endabstract]

\markboth{J.A. Peacock}{Models for large-scale structure}

\small

\sec{Background}

This conference takes place at a time when the subject of
large-scale structure is reaching a certain maturity,
following a period of very rapid development.
To appreciate how far we have come, it is instructive
to look back nearly 20 years to the 1979 Les Houches
summer school on Physical Cosmology, whose proceedings
were greatly influential in their time.

In 1979, there was an appreciation that galaxies
displayed roughly power-law correlations (measured
on small scales only), and that these would only grow
by gravity if there were some significant form of
initial fluctuations. However, the central problem of the
origin of these
initial fluctuations was scarcely mentioned. 
The only candidate primordial
spectra mentioned were expressed in terms of the
density fluctuation in a region containing $n$
particles:
\japeqn
\eqalign{
{\delta n\over n} &= n^{-1/2}\quad {\rm Poisson} \cr
&= n^{-5/6} \quad {\rm Discreteness} \cr
}
\japeqn
The former comes from random placement of particles;
the latter from random displacement of particles
within the cells of a uniform mesh. It was known
that the first gave too large an amplitude, 
so that the initial conditions had to be sub-random,
but the particle-in-box spectrum has too small
an amplitude.
There was a guess that
something like the Zeldovich spectrum was needed
($\delta n/n = 10^{-4} n^{-2/3}$), but there were
no ideas on where it might come from.

\japfig{0}{0}{519}{182}{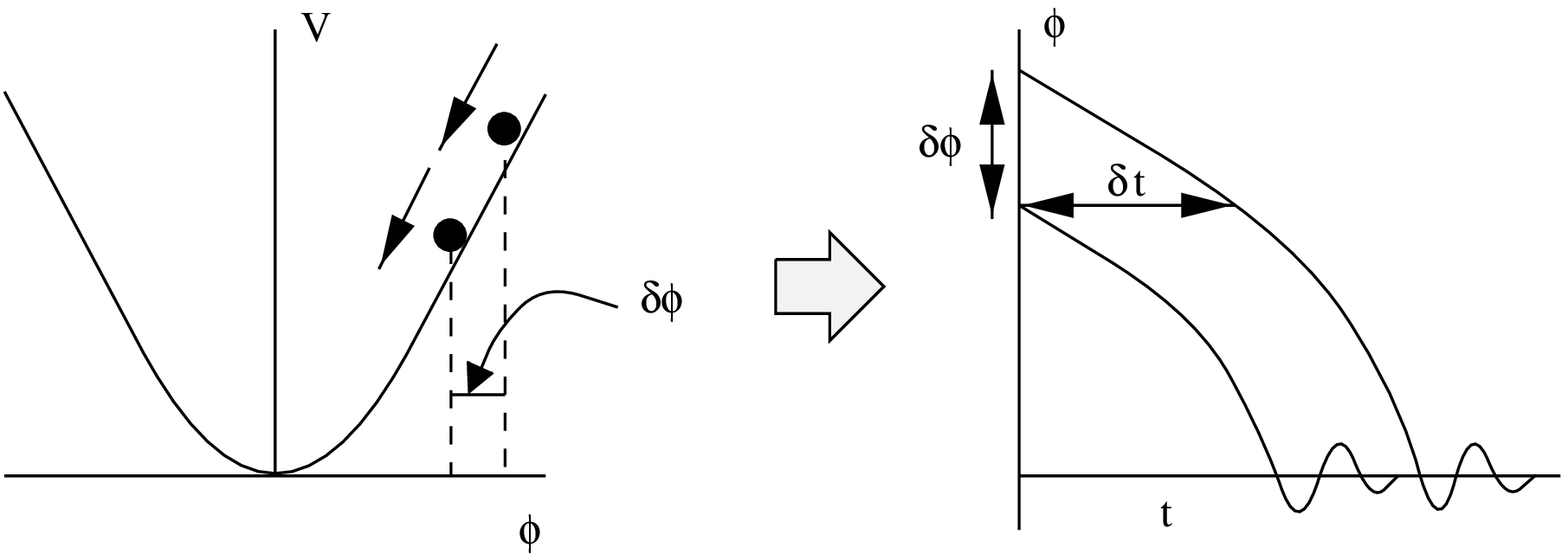}
{Inflation claims that fluctuations are seeded by
quantum fluctuations in the rolling of a scalar
field, $\phi$, down a potential, $V(\phi)$.}

The situation today is much more healthy; 
16 years have passed since it was realized that inflation
could seed these initial inhomogeneities,
and a variety of models of inflationary
structure formation have been analyzed in detail.
Furthermore, an alternative mechanism for seeding
perturbations exists through the action of
topological defects.

In its simplest form (see e.g. Liddle \& Lyth 1993; Lyth \& Riotto 1998), 
inflation deals with a single
quantum scalar field, whose expectation value
rolls down a potential (figure 1).
The equation of motion is
\japeqn
\ddot\phi+3H\dot\phi-\nabla^2\phi+{dV(\phi)\over d\phi}=0.
\japeqn
If the potential is flat in the sense
\japeqn
\eqalign{
\epsilon &\equiv {m_{\ss P}^2\over 16\pi} \,(V'/V)^2 \ll 1 \cr
\eta &\equiv {m_{\ss P}^2\over 8\pi} \,(V''/V) \ll 1 \cr
}
\japeqn
then vacuum-dominated `slow-roll' inflationary expansion 
can happen, leading to the following
horizon-scale amplitude for the density fluctuations:
\japeqn
\delta_{\ss H} = {H^2\over 2\pi\dot\phi},
\japeqn
where
\japeqn
\eqalign{
&H^2 = {8\pi\over 3}\, {V\over m_{\ss P}^2} \cr
&3H\dot\phi = -V'. \cr
}
\japeqn

There are a few free parameters here, but the characteristic
prediction of inflation is that there is a
background of gravity waves. There are thus both scalar
and tensor contributions to the CMB power spectrum:
\japeqn
C_\ell^{\ss S} \propto \ell^{n_{\ss S}-3},\quad\quad
C_\ell^{\ss T} \propto \ell^{n_{\ss T}-3}
\japeqn
(beware: different definitions of the tensor index exist).
The slopes of these spectra depend on the inflationary parameters:
\japeqn
\eqalign{
{\rm scalar\ tilt} &\equiv 1-n_{\ss S} \equiv -{d \ln \delta^2_{\ss H} \over d\, \ln k} \cr
   &=6\epsilon - 2\eta \cr
{\rm tensor\ tilt} &\equiv 1-n_{\ss T} = 2\epsilon \cr
}
\japeqn
and so does the relative tensor contribution, 
giving the consistency equation (Starobinsky 1985; Lidsey et al. 1997):
\japeqn
{C_\ell^{\ss T}\over C_\ell^{\ss S}} = 12\epsilon = 6(1-n_{\ss T})
\japeqn
These predictions can be violated in models with several scalar
fields, so inflation is a slightly soft target. Nevertheless, it
would be striking if something like these relation was to
be found in practice. To have both a potential mechanism
for generating the inhomogeneous universe plus anything approaching a
test of the theory is an astonishing achievement -- whether or
not the theory has anything to do with reality.

In testing a theory, it is good to have an alternative in mind,
but I will say little about the main class of alternatives,
which are topological defects. The state of play here will be
reviewed in Pen's talk, but presently looks unpromising
(e.g. Albrecht, Battye \& Robinson 1998).

\japfig{31}{183}{499}{579}{japfig2.eps}
{Transfer functions for various dark-matter models.
The scaling with $\Omega h^2$ is exact only for the
zero-baryon models; the baryon results are scaled from
the particular case $\Omega_{\ss B}=1$, $h=1/2$. 
}

\sec{The fluctuation spectrum}

\ssec{The linear spectrum}

The basic picture of inflationary models is therefore of a
primordial power-law spectrum, written dimensionlessly
as the logarithmic contribution to the fractional
density variance, $\sigma^2$:
\japeqn
\Delta^2(k)={d\sigma^2\over d\ln k} \propto k^{3+n},
\japeqn
where $n$ stands for $n_{\ss S}$ hereafter.
This undergoes linear growth
\japeqn
\delta_k(a) = \delta_k(a_0)\; \left[{D(a)\over D(a_0)}\right] \; T_k,
\japeqn
where the linear growth law is
\japeqn
D(a)=a\, g[\Omega(a)]
\japeqn
in the matter era,
and the growth suppression for low $\Omega$ is
\japeqn
\eqalign{
g(\Omega) &\simeq \Omega^{0.65}\ {\rm (open)} \cr
&\simeq \Omega^{0.23}\ {\rm (flat)} \cr
}
\japeqn
The transfer function $T_k$ depends on the dark-matter
content as shown in figure 2.

Note the baryonic oscillations in figure 2; these
can be significant even in CDM-dominated models
when working with high-precision data.
Eisenstein \& Hu (1998) are to be congratulated for
their impressive persistence in finding an accurate
fitting formula that describes these wiggles.
This is invaluable for carrying out a search of
a large parameter space.

\japfig{0}{0}{519}{582}{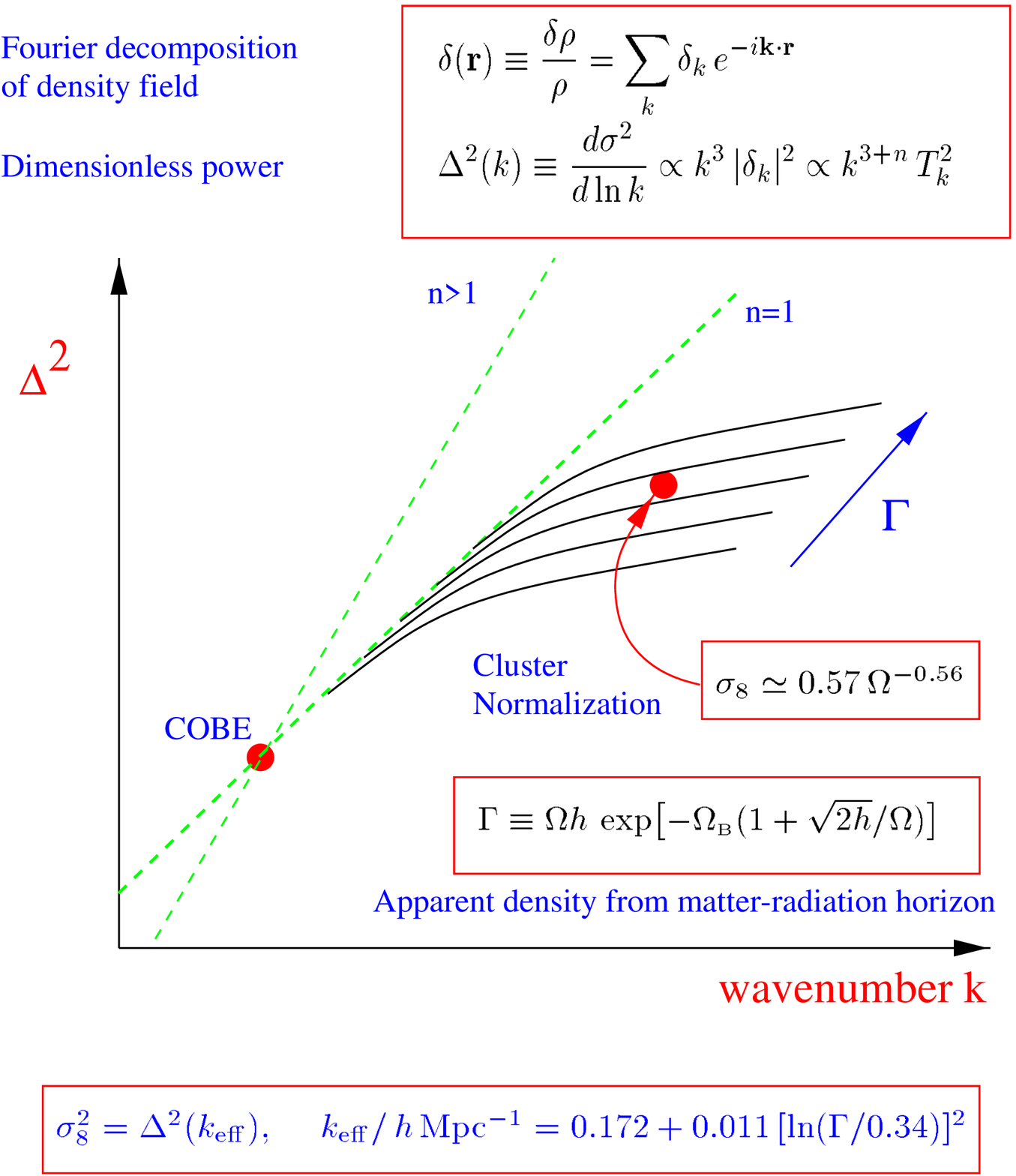}
{This figure
illustrates how the primordial power spectrum is
modified as a function of density in a CDM
model. For a given tilt, it is always
possible to choose a density that satisfies both the
COBE and cluster normalizations.}

\japfig{36}{192}{482}{578}{japfig4.eps}
{For 10\% baryons, the value of $n$ needed
to reconcile COBE and the cluster normalization
in CDM models.}

The state of the linear-theory spectrum after these
modifications is illustrated in figure 3.
The primordial power-law spectrum is reduced
at large $k$, by an amount that depends on
both the quantity of dark matter and its
nature. Generally the bend in the spectrum
occurs near $1/k$ of order the horizon size
at matter-radiation equality, $\propto 
(\Omega h^2)^{-1}$. For a pure CDM universe,
with scale-invariant initial fluctuations
($n=1$), the observed spectrum depends only on two
parameters. One is the shape $\Gamma = \Omega h$,
and the other is a normalization. On the
shape front, a government health warning is needed,
as follows. It has been quite common to take
$\Gamma$-based fits to observations as indicating
a {\it measurement\/}  of $\Omega h$, but there are
three reasons why this may give incorrect answers:

(1) The dark matter may not be CDM. An admixture of
HDM will damp the spectrum more, mimicking a
lower CDM density.

(2) Even in a CDM-dominated universe, baryons can
have a significant effect, making $\Gamma$ lower
than $\Omega h$. An approximate formula for this
is given in figure 3 (Peacock \& Dodds 1994;
Sugiyama 1995).

(3) The strongest (and most-ignored) effect is
tilt: if $n\ne 1$, then even in a pure CDM universe
a $\Gamma$-model fit to the spectrum will give a
badly incorrect estimate of the density
(the change in $\Omega h$ is roughly $0.3(n-1)$;
Peacock \& Dodds 1994).

\japfig{25}{180}{498}{599}{japfig5.eps}
{The velocity power spectrum, according to linear theory.
The two solid curves show CDM models with $\Omega=1$
and $0.3$ (flat), with $h=0.65$ and the appropriate value
of $n$ chosen as in figure 4 in order to satisfy both
the cluster-abundance $\sigma_8$ and the COBE
normalization.
The points show the APM power spectrum linearized as
described in Peacock (1997), for the same two
cosmologies, and also with the cluster normalization.
All these determinations are in general agreement over
the amplitude and broad range of the velocity spectrum.
}

\ssec{Normalization}

The other parameter is the normalization.
This can be set at a number of points.
The COBE normalization comes from large-angle
CMB anisotropies, and is sensitive to the
power spectrum at $k\simeq 10^{-3}\hompc$.
The alternative is to set the normalization
near the quasilinear scale, using the abundance of
rich clusters. Many authors have tried this
calculation, and there is good agreement on the
answer:
\japeqn
\sigma_8 \simeq (0.5 - 0.6) \, \Omega_m^{-0.6}.
\japeqn
(White, Efstathiou \& Frenk 1993; Eke et al. 1996; Viana \& Liddle 1996).
In many ways, this is the most sensible normalization
to use for LSS studies, since it does not rely
on an extrapolation from larger scales.

Within the CDM model, it is always possible to satisfy
both these normalization constraints, by appropriate
choice of $\Gamma$ and $n$. This is illustrated in
figure 4. Note that vacuum energy affects the answer;
for reasonable values of $h$ and reasonable
baryon content, flat models require $\Omega_m\simeq 0.3$, whereas
open models require $\Omega_m\simeq 0.5$.

\ssec{The velocity spectrum}

Given the density spectrum, other
quantities of interest can be calculated in linear
theory. For example, the velocity spectrum is
\japeqn
{d\sigma_v^2\over d\ln k} = \left({\Omega_m^{0.6}H\over k}\right)^2\Delta^2(k).
\japeqn
A plot of this function is shown in figure 5,
determined in various ways. If these determinations are all
normalized according to the cluster abundance,
there appears to be relatively little uncertainty
in the velocity spectrum. This is not surprising, since the
cluster normalization is tied to the observed abundance
of a set of objects of a fixed observed velocity
dispersion.

Note the impressive breadth of the velocity spectrum:
even modes of $600\mpcoh$ wavelength contribute
about $100\kms$ rms to the motion of the local
group. This shows why it has been hard to obtain
redshift surveys of sufficient depth to obtain a convergent
prediction of the local gravitational dipole.

These velocity fields are responsible for the
distortion of the clustering pattern in 
redshift space, as first clearly articulated
by Kaiser (1987). 
For a survey that subtends a small angle
(i.e. in the distant-observer approximation), a good approximation to
the anisotropic redshift-space Fourier spectrum is given
by the Kaiser function together with a damping
term from nonlinear effects:
\japeqn
\delta_k^s=\delta_k^r (1+\beta\mu^2)D(k\sigma\mu),
\japeqn
where $\beta=\Omega_m^{0.6}/b$, $b$ being the
linear bias parameter of the galaxies under study,
$\mu={\bf \hat k \cdot \hat r}$, and 
$D(y)\simeq (1+y^2/2)^{-1/2}$, and $\sigma\simeq 350\kms$
is the pairwise velocity dispersion of galaxies
(e.g. Ballinger, Peacock \& Heavens 1996).
In principle, this distortion should be a robust
way to determine $\Omega$ (or at least $\beta$);
see Strauss \& Willick (1995) and Hamilton (1997)
for details of the practical application of these
ideas.

\ssec{The nonlinear spectrum}

On smaller scales ($k\gs 0.1$), nonlinear effects become important.
These are relatively well understood so far as they affect the
power spectrum of the mass (e.g. 
Hamilton et al. 1991; Jain, Mo \& White 1995;
Peacock \& Dodds 1996). These methods were applied
by Peacock \& Dodds (1994; PD94) to estimate the linear
spectrum from the observed spectrum of a number of tracers
(figure 6). To within a scatter of perhaps a factor 1.5 in
power, the results were consistent with a $\Gamma\simeq 0.25$ CDM model.
Even though the subsequent sections will discuss some possible
disagreements with the CDM models at a higher level of
precision, the general existence of CDM-like curvature
in the spectrum is likely to be an important clue to the
nature of the dark matter.

\japfig{37}{117}{490}{749}{japfig6.eps}
{The PD94 compilation of power-spectrum
measurements. The upper panel shows raw
power measurements; the lower shows these data
corrected for relative bias, nonlinear effects, and
redshift-space effects.}

One interesting aspect of the existing spectrum
determinations is that they are relatively smooth,
whereas figure 2 shows that rather large oscillatory
features would be expected if the universe was baryon
dominated. The lack of such features is one reason
for believing that the universe might be dominated
by collisionless nonbaryonic matter (consistent with
primordial nucleosynthesis if $\Omega_m\gs 0.1$).

\japfig{0}{25}{487}{544}{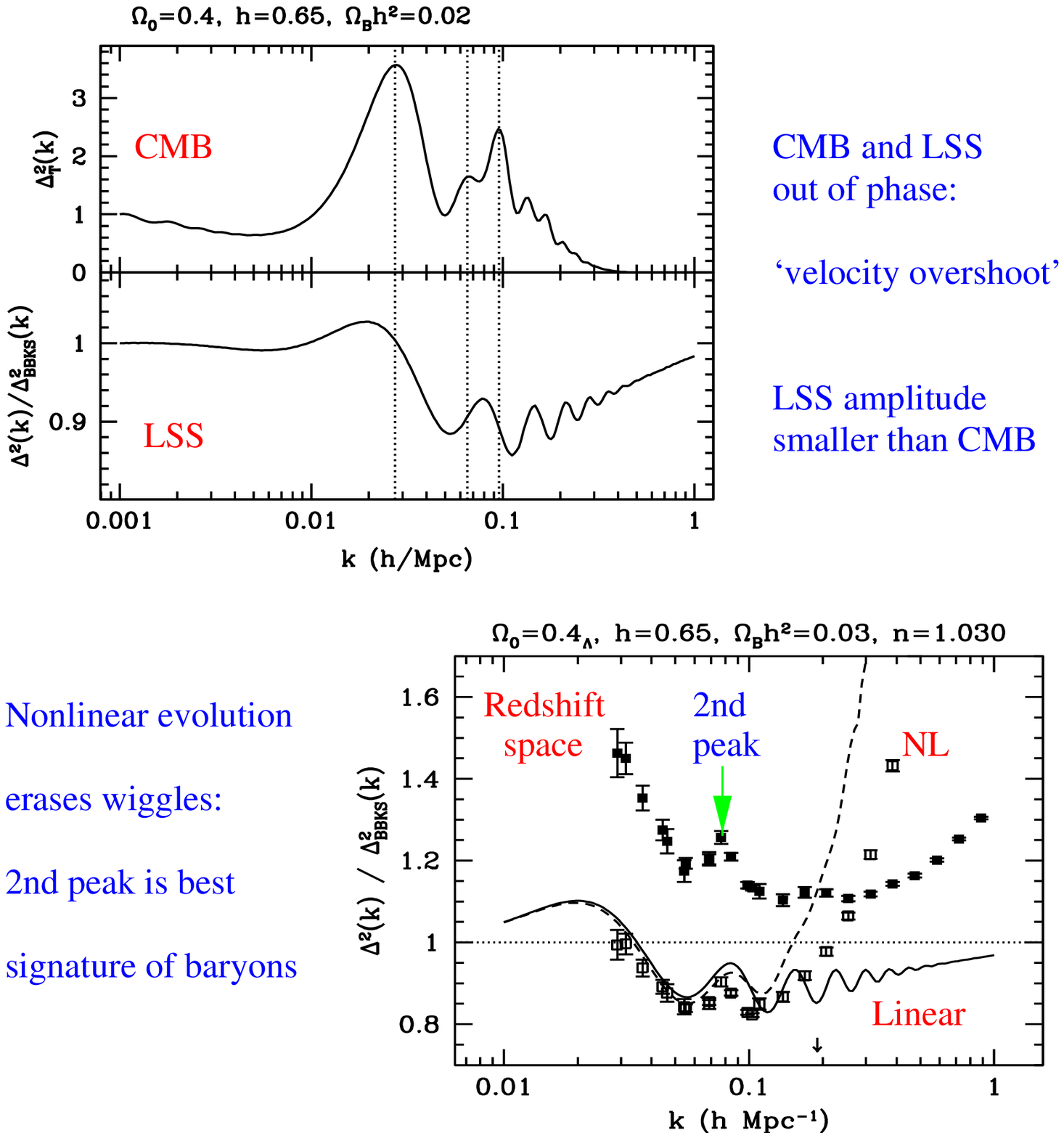}
{Baryonic fluctuations in the spectrum can
become significant for high-precision measurements.
Although such features are much less important in the
density spectrum than in the CMB (first panel), the
order 10\% modulation of the power is potentially
detectable. However, nonlinear evolution has the
effect of damping all beyond the second
peak. This second feature is relatively
narrow, and can serve as a clear proof of the past
existence of oscillations in the baryon-photon fluid
(Meiksin, White \& Peacock 1998).}

Nevertheless, 
baryonic fluctuations in the spectrum can
become significant for high-precision measurements.
Figure 7 shows that order 10\% modulation of the
power may be expected in realistic baryonic
models (Eisenstein \& Hu 1998; Goldberg \& Strauss 1998).
Most of these features are however removed
by nonlinear evolution.
The highest-$k$ feature to survive is usually
the second peak, which almost always lies
near $k=0.05\,{\rm Mpc}^{-1}$ (no $h$, for a change).
This feature is relatively
narrow, and can serve as a clear proof of the past
existence of baryonic oscillations in forming the 
mass distribution
(Meiksin, White \& Peacock 1998).
However, figure 7 emphasises that the easiest way
of detecting the presence of baryons is likely to
be through the CMB spectrum. The oscillations
have a much larger `visibility' there, because
the small-scale CMB anisotropies come directly
from the coupled radiation-baryon fluid, rather
than the small-scale dark matter perturbations.

\ssec{The APM spectrum}

In the past few years, much attention has been attracted by
the estimate of the galaxy power spectrum from the APM
survey (Baugh \& Efstathiou 1993, 1994; Maddox et al. 1997).
The APM result was generated from
a catalogue of $\sim 10^6$ galaxies; because it is based
on a deprojection of angular clustering, it is immune to the
complicating effects of redshift-space distortions.

\japfig{20}{75}{525}{700}{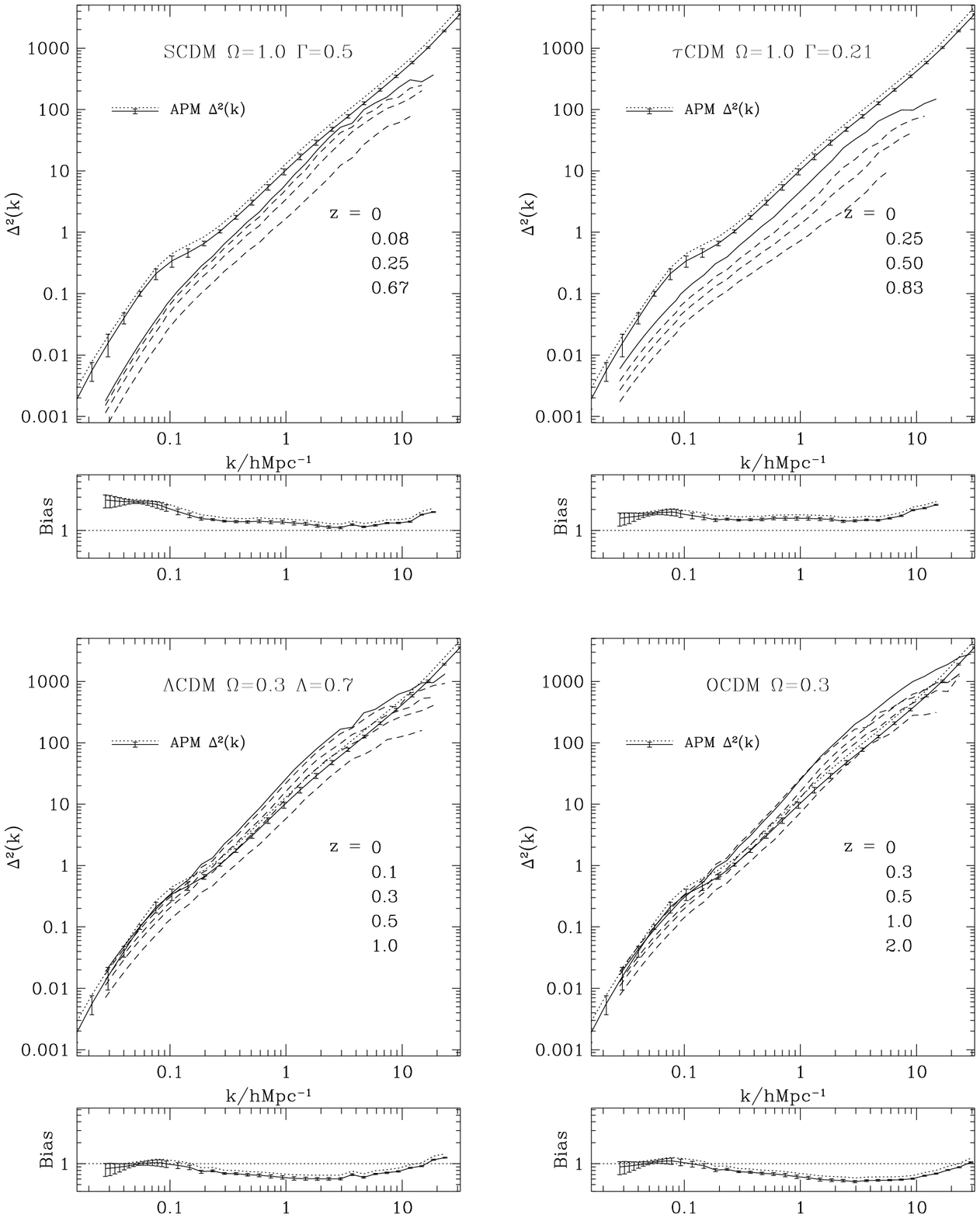}
{The nonlinear evolution of various CDM
power spectra, as determined by the Virgo
consortium (Jenkins et al. 1998).}

The error estimates are derived empirically from the scatter
between independent regions of the sky, and so should be
realistic. If there are no undetected systematics, these
error bars say that the power is very accurately determined,
and this has significant implications if true.

A number of authors have pointed out that the detailed
spectral shape appears to be inconsistent with that of nonlinear
evolution from CDM initial conditions.
(e.g. Efstathiou,  Sutherland \& Maddox 1990;
Klypin, Primack \& Holtzman 1996; Peacock 1997).
Perhaps the most detailed work was carried out by the
VIRGO consortium, who carried out $N=256^3$ simulations of
a number of CDM models (Jenkins et al. 1998). Their results
are shown in figure 8, which gives the nonlinear power
spectrum at various times (cluster normalization is chosen
for $z=0$) and contrasts this with the APM data.
The lower small panels are the scale-dependent bias
that would  required if the model did in fact describe the real universe,
defined as
\japeqn
b(k)\equiv \left({\Delta^2_{\rm gals}(k)\over\Delta^2_{\rm mass}}\right)^{1/2}.
\japeqn
In all cases, the required bias is non-monotonic; it rises at
$k\gs 5\mpcoh$, but also displays a bump around
$k\simeq 0.1\mpcoh$.
If real, this feature seems impossible to understand as a
genuine feature of the mass power spectrum; certainly, it is not at
a scale where the effects of even a large baryon fraction
would be expected to act (Eisenstein et al. 1998; Meiksin, White
\& Peacock 1998).

\japfig{30}{190}{488}{589}{japfig9.eps}
{a comparison of the linearized APM power spectrum with
two CDM models that satisfy COBE and
cluster normalization.}

An alternative way of presenting this difficulty for
CDM models is shown in figure 9. Here, the methods of
Peacock \& Dodds (1996) have been used in an attempt to
recover the empirical linear spectrum from the APM data.
It is assumed that the bias is independent of scale, and
that the cluster normalization applies.
The results (for low and high $\Omega$) are contrasted with 
CDM models with reasonable Hubble parameters and baryon
fractions. Interestingly, there is not a huge difference
in shape between the CDM models. This arises because both
are also consistent with COBE, and therefore have very
different values of the tilt, as above. This plot should drive
home the lesson that the apparent value of $\Gamma$ cannot
be used as an estimate of $\Omega h$ without further,
possibly fragile, assumptions.

In any case, the linear spectrum that would evolve into the
APM result is seen in figure 9 to differ significantly from the
CDM models. The bump at $k\simeq 0.1\mpcoh$ is apparent,
as is a flattening at larger $k$ -- the effective
spectral index appears to approach $n=-3$, whereas the CDM
models are never this flat. 

A spectrum of this form
would be expected with a roughly 30\% admixture
of massive neutrinos in a total $\Omega \simeq 1$
(Klypin et al. 1993; Pogosyan \& Starobinsky 1995).
This success in matching the empirical spectrum shape
is a very significant attraction of the MDM model.
However, there are also very serious difficulties with
the idea of an MDM universe, related to the formation
of high-redshift objects. MDM suffers from the same
difficulties as any $\Omega=1$ model in making high-redshift
massive clusters (e.g. Henry 1997). This alone may be a strong enough
argument to reject the model. However, the very flatness
of the high-$k$ spectrum that is attractive in terms of
clustering statistics makes it difficult for MDM to
assemble the observed high-redshift galaxies
(e.g. Mo \& Miralda-Escud\'e 1994; Mo \& Fukugita 1996;
Peacock et al. 1998). These problems are eased in
a low-density model, which is probably
required by the supernova Hubble diagram
(Perlmutter et al 1997; Riess et al. 1998), but the MDM spectral shape is
then wrong.

\sec{Towards an understanding of bias}

The conclusions from the above discussion are either that
the physics of dark matter and structure formation are
more complex than in CDM models, or that the relation
between galaxies and the overall matter distribution
is sufficiently  complicated that the effective bias is
not a simple slowly-varying monotonic function of position.

\ssec{Local bias}

The simplest assumption is that all the complicated physical effects 
leading to galaxy formation depend in a causal (but nonlinear) way on the
local mass density, so that we write
\japeqn
\rho_{\rm light}=f(\rho_{\rm mass}).
\japeqn
Coles (1993) showed that, under rather general assumptions, this
equation would lead to an effective bias that was a monotonic
function of scale. This issue was investigated in some detail
by Mann, Peacock \& Heavens (1998), who verified Coles'
conclusion in practice for simple few-parameter forms for
$f$, and found in all cases that the effective bias varied
rather weakly with scale. The APM results thus are either inconsistent
with a CDM universe, or require non-local bias.

\japfig{0}{0}{484}{472}{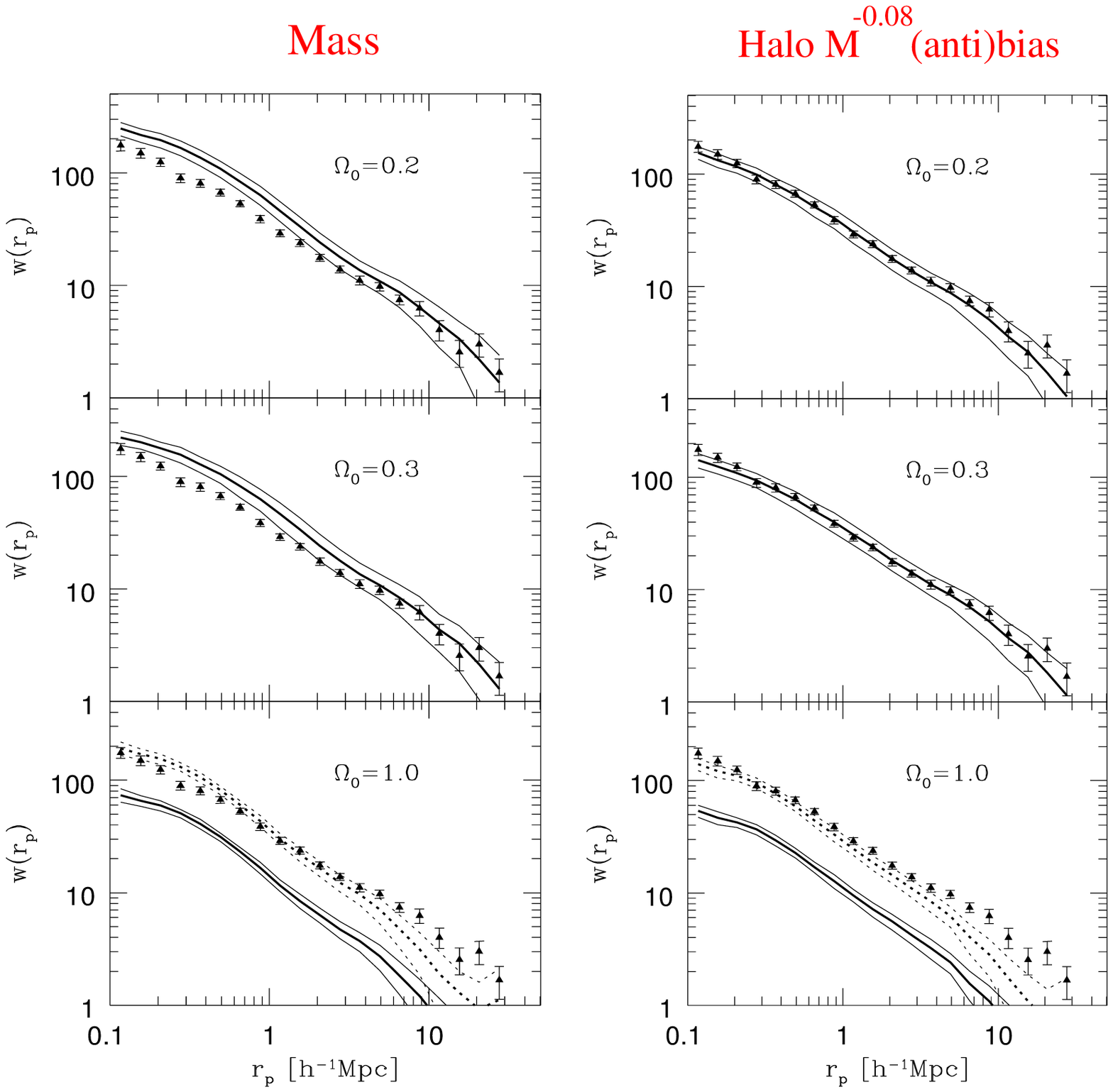}
{The projected correlation function from the
LCRS fails to match CDM models when comparison is made to
just the mass distribution. However, the agreement
is excellent when allowance is made for a small
degree of scale-dependent antibias; galaxy formation
is suppressed in the most massive haloes
(Jing, Mo \& B\"orner 1998).}

A puzzle with regard to this conclusion is provided
by the work of Jing, Mo \& B\"orner (1998). They
evaluated the projected real-space correlations
for the LCRS survey (see figure 10). This statistic also
fails to match the prediction of CDM models, but this
can be amended by introducing a simple {\it antibias\/}
scheme, in which galaxy formation is suppressed in the
most massive haloes. This scheme should in practice
be very similar to the Mann, Peacock \& Heavens recipe
of a simple weighting of particles as a function of the
local density; indeed, the main effect is a change of
amplitude, rather than shape of the correlations.
The puzzle is this: if the APM power spectrum is used
to predict the projected correlation function, the
result agrees almost exactly with the LCRS. Either
projected correlations are a rather
insensitive statistic, or perhaps the Baugh \& Efstathiou
deconvolution procedure used to get $P(k)$ has
exaggerated the significance of features in the spectrum.
The LCRS results are one reason for treating the apparent conflict 
between APM and CDM with caution.

\ssec{High-order correlations}

A further piece of evidence concerning bias comes from the
higher-order correlations of the density field, which are
most conveniently described via the hierarchical moments:
\japeqn
S_J \equiv {\langle \delta^J \rangle / [\langle \delta^2 \rangle ]^{J-1}}.
\japeqn
These are defined so that, for gravitationally-driven growth of
fluctuations, the $S_J$ are dimensionless numbers that can be calculated
in perturbation theory. For example, if the density field is smoothed
with a spherical top-hat window, the third moment for a Gaussian field
should be
\japeqn
S_3 ={34\over 7} - (3+n)
\japeqn
(Bernardeau 1994). In the presence of pure linear bias of the density field,
$\delta_{\rm gals}=b\delta_{\rm mass}$, $S_J\rightarrow b^{2-J} S_J$,
so there might seem to be an interesting test here both of Gaussianity and
of whether light traces mass. In practice, 
Gazta\~naga \& Frieman (1994) found that higher-order results from
(again) the APM catalogue were in remarkably good agreement with
the simplest unbiased Gaussian predictions. However, the interpretation
of this result needs care: linear density bias is unphysical, in that
it can yield $\delta<-1$. Gravitational evolution causes
increasing skewness to develop because this is the only way in which
a large second moment $\langle\delta^2\rangle \gg 1$ can coexist
with the $\delta>-1$ constraint. Any bias scheme that has the
effect of increasing $\langle\delta_{\rm gals}^2\rangle$
relative to $\langle\delta_{\rm mass}^2\rangle$ is almost certain to
increase the skewness also. The higher-order results are thus
quite a significant limit on non-Gaussianity, but 
do not strongly constrain bias.

\japfig{0}{0}{429}{481}{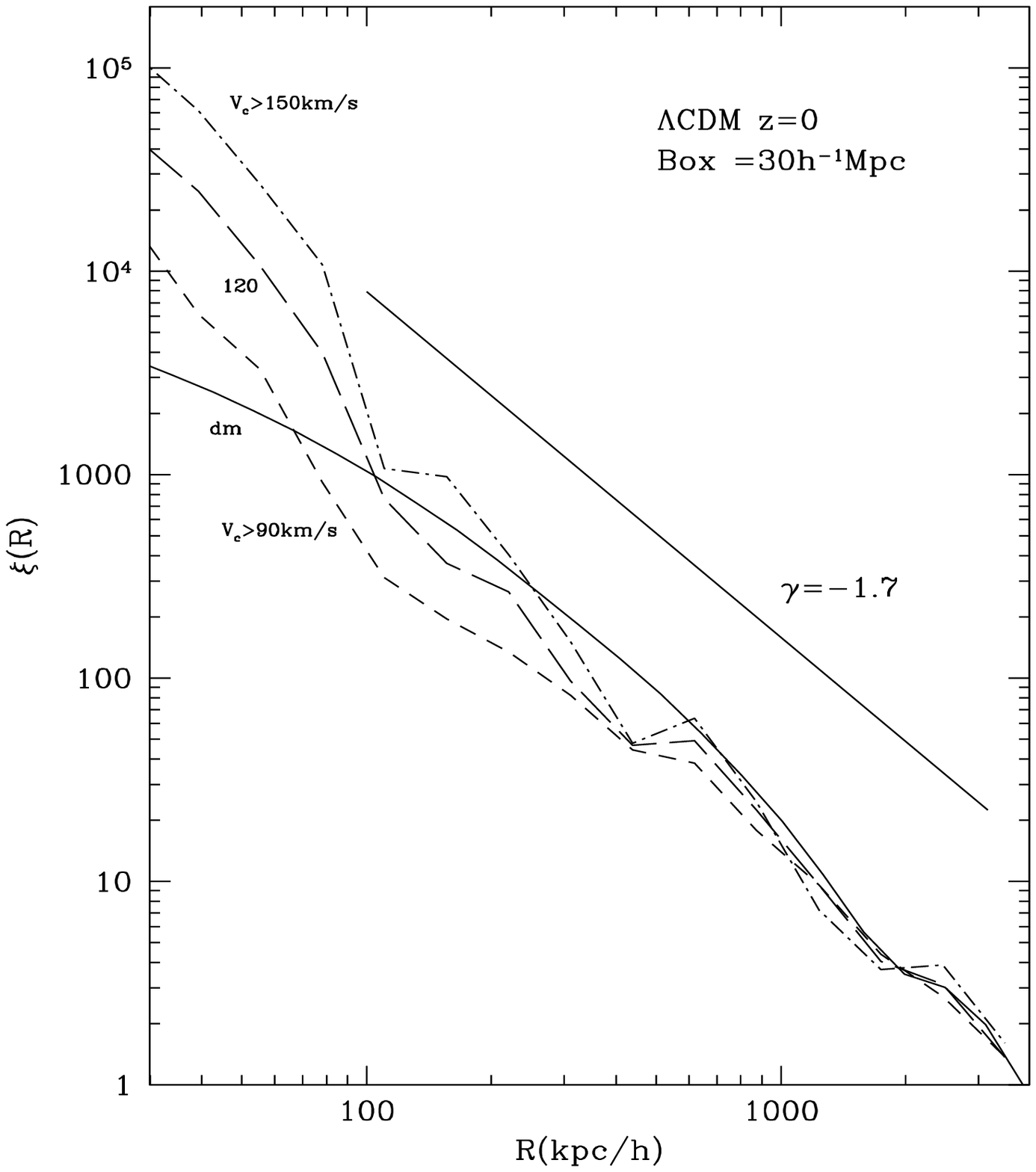}
{Klypin et al. (1997) have shown that the correlation function of
galaxy-scale haloes is much closer to a single power law than
the correlation function of the overall mass distribution.}

\ssec{Halo correlations}

In reality, bias is unlikely to be completely causal,
and this has led some workers to explore stochastic bias
models, in which
\japeqn
\rho_{\rm light}=f(\rho_{\rm mass}) + \epsilon,
\japeqn
where $\epsilon$ is a random field that is uncorrelated with the
mass density (Pen 1998; Dekel \& Lahav 1998).
Although truly stochastic effects are possible in galaxy formation,
a relation of the above form is expected when the
galaxy and mass densities are filtered on some scale
(as they always are, in practice). Just averaging a 
galaxy density that is a nonlinear
function of the mass will lead to some scatter when comparing with the
averaged mass field; a scatter will also arise when the
relation between mass and light is non-local, however, and this
may be the dominant effect.

The simplest and most important example of non-locality
in the galaxy-formation process is to recognize that
galaxies will generally form where there are galaxy-scale
haloes of dark matter. In the past, it was
generally believed that dissipative processes were
critically involved in galaxy formation, since pure
collisionless evolution would lead to the destruction
of galaxy-scale haloes when they are absorbed into the
creation of a larger-scale nonlinear system such as a group
or cluster. However, it turns out that this
{\it overmerging problem\/} was only an artefact of
inadequate resolution. When a simulation is carried out
with $\sim 10^6$ particles in a rich cluster, the cores of
galaxy-scale haloes can still be identified after many crossing
times (Ghigna et al. 1997). This is a very important result,
since it holds out the hope that many of the issues concerning
where galaxies form in the cosmic density field can be settled
within the domain of collisionless simulations. 
Dissipative physics will still be needed to understand in
detail the star-formation history within a galaxy-scale
halo. Nevertheless, the idea that there may be a one-to-one correspondence
between galaxies and galaxy-scale dark-matter haloes
is clearly an enormous simplification -- and one that increases
the chance of making robust predictions of the statistical
properties of the galaxy population.

Another piece of work which also emphasises the survival of
galaxy-scale haloes is Klypin et al. (1997; see also
Kravtsov's talk at this meeting). Rather than looking
only at a single cluster, Klypin et al. attempt to
find galaxy haloes within a random cosmological
volume. Their results are shown in figure 11, where the
correlation function for the mass in a $\Lambda$CDM
simulation is contrasted with the correlation function for
the haloes. The mass $\xi(r)$ shows the familiar form:
the sharp quasilinear rise between $\xi\simeq 10$ and
300, followed by the flattening associated with virialization.

By contrast, the halo correlation functions are much closer
to single power laws: they are antibiased at $\xi\simeq 100$,
but positively biased for $\xi\gs 1000$. These are qualitatively
the trends that are required in order to reconcile the APM data with CDM
models, so this is clearly a result of great potential significance.
Future work should improve the precision of the results and
allow the simulation boxes to be made larger (they are presently
uncomfortably small). However, if the result holds up, we will
be no nearer an {\it understanding\/} of what is going on.
A simple result like power-law correlations demands to be explained
in a more intuitive manner. If the result turns out to be
inevitable on some general grounds, then this may be bad news
for the idea of large-scale-structure studies as a diagnostic
tool -- it is possible that the galaxy correlations may turn out
to be less sensitive to the underlying linear power spectrum than
the mass correlations are.

\sec{Clustering at high redshifts}

A longstanding ambition has been to remove some of the uncertainties
concerning the generation of large-scale structure by observing
the growth of clustering with time.
After a number of years of ambiguous results from angular clustering
of faint galaxies, this has at last proved possible by using
deep redshift surveys. 

Perhaps the first convincing measurement of clustering evolution
came from the CFRS (Le F\`evre et al. 1996), which indicated evolution
in the sense that $\xi$ at a given comoving separation was 
a factor of about four smaller at $z=1$ than it is today.
Evolution in the same sense (but somewhat less strong) has
been found by the CNOC2 team (Carlberg et al. 1998).

These results appear to prove that clustering does indeed grow
by gravity, but this idea looks less convincingly demonstrated when the
results from clustering at higher redshifts are
considered. It is remarkable that samples at $z\simeq 3.5$
are now as large as those at $0.5\ls z \ls 1$ were only two years ago;
this shows how efficient the Lyman-break selection technique has been.
At this highest redshift, we have the remarkable result that
the comoving correlation strength appears to be very
close to that measured today
(Steidel et al. 1998; Adelberger et al. 1998).
In other words, any downward evolution in $\xi$
between $z=0$ and $z=1$ must be reversed between
$z=1$ and $z=3.5$, so that $\xi(z)$ has a minimum.

Alternatively, if we believe that clustering of the mass
grows monotonically according to gravitational instability,
then clustering at $z=3.5$ must be very strongly biased.
Steidel et al. (1998)
obtain a minimum bias of 6, with a preferred
value of 8, assuming $\Omega=1$.
It is easy to
translate to other models, since we observe cell variances $\sigma^2_{\rm cell}$
directly, where the cell has a given angular width and depth in redshift.
For low $\Omega$ models, the cell volume
will increase by a factor $[S_k^2(r)\, dr]/[S_k^2(r_1)\, dr_1]$; comparing with
present-day fluctuations on this larger scale will tend to increase
the bias. However, for low $\Omega$, two other effects increase the
predicted density fluctuation at $z=3$: the cluster constraint
increases the present-day fluctuation by a factor $\Omega^{-0.56}$, and
the growth between redshift 3 and the present will be less than
a factor of 4. Applying these corrections gives
\japeqn
{ b(\Omega=0.3) \over b(\Omega=1) } =
\left\{ {0.42\ ({\rm open}) \atop 0.60\ \rlap{({\rm flat})}
\phantom{({\rm open})}} \right. ,
\japeqn
which suggests an approximate scaling as $b\propto \Omega^{0.72}$ (open)
or $\Omega^{0.42}$ (flat). Strong bias is needed at $z=3$ for all reasonable
values of $\Omega$.

The standard explanation for this effect is high-peak
bias, which is bound to operate if the high-redshift
galaxies are rare, newly-forming systems.
The linear bias parameter depends on the rareness of
the fluctuation and the rms of the underlying field as
\japeqn
b=1+{\nu^2-1\over \nu\sigma}= 1+ {\nu^2-1\over \delta_c}
\japeqn
(Kaiser 1984; Cole \& Kaiser 1989; Mo \& White 1996 -- but see
also Jing 1998),
where $\nu = \delta_c/\sigma$, and $\sigma^2$ is the
fractional mass variance at the redshift of interest.
A bias at the observed level is therefore to
be expected, provided the Lyman-break galaxies are moderately massive
systems, with circular velocities $V\simeq 150\kms$
(e.g. Bagla 1998a, b; Peacock et al. 1998).
The minimum in $\xi$ makes sense since we observe
something proportional to $(b\sigma)^2$;
$b$ is constant for large $\sigma$, but scales as $\sigma^{-2}$
for small $\sigma$; the observed clustering therefore has a
minimum at the redshift where $\sigma\simeq 1$.

In order to turn this plausibility argument into a proof, two
tests should be carried out. First, the circular velocities of
Lyman-break galaxies need to be measured; second, even larger and
more detailed
redshift surveys need to be carried out at $z\simeq 3$, in
order to verify that the structures causing $\xi(r)$ have the
same topological character of voids and filaments as seen
in local large-scale structure. If these criteria are
satisfied, then we can claim to have some understanding of
the growth of structure, and the existing clustering
results may be used as a test of models.

\sec{Conclusions}

It should be clear from this talk that large-scale structure has 
advanced enormously as a field in the past two decades.
Many of our long-standing ambitions have been realised;
in some cases, much faster than we might have expected.
Of course, solutions for old problems generate new difficulties.
Although we now have good measurements of the clustering
spectrum and its evolution, it is less clear that the
placement of galaxies with respect to the mass is simple
enough to use these results as direct statistics with which
to test theories. A fairly safe bet is that one of the
major results from new large surveys such as 2dF and
Sloan will be a heightened appreciation of the subtleties
of this problem.

Nevertheless, we should not be depressed that problems
remain. Observationally, we are moving from an era of
20\% -- 50\% accuracy in measures of large-scale
structure to a future of pinpoint precision. This
maturing of the subject will demand more careful
analysis and rejection of some of our existing tools
and habits of working. The prize for rising to this
challenge will be the ability to claim a real understanding
of the origin of structure in the universe. We are
not there yet, but there is a real prospect that the next
5--10 years may see this remarkable goal achieved.

\section*{References}

\begin{thedemobiblio}{}
\japref Adelberger K., Steidel C., Giavalisco M., Dickinson M., Pettini M., Kellogg M., 1998, ApJ, 505, 18
\japref Albrecht A., Battye R.A., Robinson J., 1997, astro-ph/9711121
\japref Bagla J.S., 1998b, MNRAS, 299, 417
\japref Bagla J.S., 1998a, MNRAS, 297, 251
\japref Ballinger W.E., Peacock J.A., Heavens A.F., 1996, MNRAS, 282, 877
\japref Baugh C.M., Efstathiou G., 1993, MNRAS, 265, 145
\japref Baugh C.M., Efstathiou G., 1994, MNRAS, 267, 323
\japref Bernardeau F., {1994}, {\apj}, {433}, {1}
\japref Carlberg R.G. et al., 1998, astro-ph/9805131
\japref Cole S., Kaiser N., 1989, MNRAS, 237, 1127
\japref Coles P., {1993}, {MNRAS}, {262}, {1065}
\japref Dekel A., Lahav O., 1998, astro-ph/9806193
\japref Efstathiou G., Sutherland W., Maddox S.J., 1990,  Nature, 348, 705
\japref Eisenstein D.J., Hu W., 1998, ApJ, 496, 605
\japref Eisenstein D.J., Hu W., Silk J., Szalay A.S., 1998, ApJ, 494, L1
\japref Eke V.R., Cole S., Frenk C.S., {1996}, {MNRAS}, {282}, {263}
\japref Gazta\~naga E., Frieman J.A., 1994, ApJ, 437, 13
\japref Ghigna S., Moore B., Governato F., Lake G., Quinn T., Stadel J., 1998, MNRAS, 300, 146
\japref Goldberg D.M., Strauss M.A., 1998, ApJ, 495, 29
\japref Hamilton A.J.S., Kumar P., Lu E.,  Matthews A., 1991, \apj, 374, L1 
\japref Hamilton A.J.S., 1997, astro-ph/9708102
\japref Henry J.P., 1997, ApJ, 489, L1
\japref Jain B., Mo H.J., White S.D.M., 1995, \mn, 276, L25 
\japref Jenkins A., Frenk C.S., Pearce F.R., Thomas P.A., Colberg J.M., White S.D.M., Couchman H.M.P., Peacock J.A., Efstathiou G., Nelson A.H., 1998, ApJ, 499, 20
\japref Jing Y.P., 1998, ApJ, 503, L9
\japref Jing Y.P., Mo H.J., B\"orner G., {1998}, ApJ, 494, 1
\japref Kaiser N., 1984, ApJ, {284}, L9
\japref Kaiser N., 1987, MNRAS, 227, 1
\japref Klypin A., Holtzman J., Primack J., Reg\H os E., {1993}, {\apj}, {416}, {1}
\japref Klypin A., Primack J., Holtzman J., 1996, \apj, 466, 13
\japref Klypin A., Gottloeber S., Kravtsov A.V., Khokhlov A.M., 1997, astro-ph/9708191
\japref Le F\`evre O., et al., 1996. ApJ, 461, 534
\japref Liddle A.R., Lyth D., {1993}, {Phys. Reports}, {231}, {1}
\japref Lidsey J.E. et al., 1997, Rev. Mod. Phys., 69, 373
\japref Lyth D.H., Riotto A., 1998, hep-ph/9807278
\japref Maddox S. Efstathiou G., Sutherland W.J., 1996, MNRAS, 283, 1227
\japref Mann R.G., Peacock J.A., Heavens A.F., {1998}, MNRAS, 293, 209
\japref Meiksin A.A., White M., Peacock J.A., MNRAS, submitted
\japref Mo H.J., Miralda-Escud\'{e} J., 1994, {\apj}, {430}, L25
\japref Mo H.J., Fukugita M., 1996, {\apj}, {467}, L9
\japref Mo H.J., White S.D.M., 1996, MNRAS, 282, 1096
\japref Peacock J.A., Dodds S.J., 1994, MNRAS, 267, 1020
\japref Peacock J.A., Dodds S.J., 1996, MNRAS, 280, L19
\japref Peacock J.A., 1997, \mn, 284, 885
\japref Peacock J.A., Jimenez R., Dunlop J.S., Waddington I., Spinrad H., Stern D., Dey A., Windhorst R.A., 1998, MNRAS, 296, 1089
\japref Pen, W.-L., 1998, ApJ, 504, 601
\japref Perlmutter S. et al., 1997, ApJ, 483, 565
\japref Pogosyan D.Y., Starobinsky A.A., 1995, ApJ, 447, 465
\japref Riess A.G. et al., 1998, A.J., 116, 1009
\japref Starobinsky A.A., 1985, Sov. Astr. Lett., 11, 133
\japref Steidel C.C., Adelberger K.L., Dickinson M., Giavalisco M., Pettini M., Kellogg M., 1998, ApJ, 492, 428
\japref Strauss M.A., Willick J.A., 1995, Physics Reports, 261, 271
\japref Sugiyama N., {1995}, {\apjs}, {100}, {281}
\japref Viana P.T., Liddle A.R., 1996, MNRAS, 281, 323
\japref White S.D.M., Efstathiou G., Frenk C.S., 1993, {\mn}, {262}, 1023
\end{thedemobiblio}{}
\end{document}